# Identifying a severity measure for head acceleration events associated with suspected concussions

Gregory Tierney[1], Ross Tucker[2,3], James Tooby[4], Lindsay Starling [2,5], Éanna Falvey[2,6], Danielle Salmon[2], James Brown[3,4], Sam Hudson[5], Keith Stokes[5,7], Ben Jones,[4,8,9,10,11], Simon Kemp[7,12], Patrick O'Halloran[5,13], Matt Cross[8], Melanie Bussey[14], David Allan[1]

[1] Nanotechnology and Integrated Bioengineering Centre (NIBEC), School of Engineering, Ulster University, Belfast, United Kingdom

[2] World Rugby, 8-10 Pembroke St., Dublin, Ireland

[3] Institute of Sport and Exercise Medicine, Stellenbosch University, South Africa

[4] Carnegie Applied Rugby Research (CARR) Centre, Carnegie School of Sport, Leeds Beckett University, Leeds, United Kingdom

[5] UK Collaborating Centre on Injury and Illness Prevention in Sport (UKCCIIS), University of Bath, United Kingdom

[6] School of Medicine & Health, University College Cork, Cork, Ireland

[7] Rugby Football Union, Twickenham, United Kingdom

[8] Premiership Rugby, London, United Kingdom

[9] England Performance Unit, Rugby Football League, Manchester, United Kingdom

[10] School of Behavioural and Health Sciences, Faculty of Health Sciences, Australian Catholic University, Brisbane, QLD, Australia

[11] Division of Physiological Sciences and Health through Physical Activity, Lifestyle and Sport Research Centre, Department of Human Biology, Faculty of Health Sciences, University of Cape Town, Cape Town, South Africa

[12] London School of Hygiene and Tropical Medicine, London, United Kingdom

[13] Marker Diagnostics UK Ltd, United Kingdom

[14] School of Physical Education Sport and Exercise Sciences, University of Otago, Dunedin, New Zealand

**Corresponding author:** Dr Gregory Tierney, Ulster University, Belfast, United Kingdom.

Email: g.tierney@ulster.ac.uk

**Contributors:** GT conceptualised the research project. All authors were involved in the design and data collection for the study. GT and DA were responsible for the analysis and interpretation of the results. GT and DA drafted the manuscript. All authors critically reviewed and edited the manuscript prior to submission.

**Competing Interests:** GT and BJ have received research funding from Prevent Biometrics and World Rugby. KS and MB have received research funding from World Rugby. LS, RT, EF, DS, JB are employed by or contracted as consultants to World Rugby. GT previously conducted consultancy work for World Rugby. KS and SK are employed by the Rugby Football Union. SH receives funding for his PhD studies from the Rugby Football Union and Premiership Rugby. BJ is a consultant with Premiership Rugby and the Rugby Football League. MC is employed by Premiership Rugby and was previously employed by the Rugby Football Union. POH has previously been contracted by the Rugby Football Union and is employed by Marker Diagnostics UK Ltd, a company developing salivary biomarker testing for sport related concussion. DA and JT declare they have no conflicts of interest.

**Funding:** Funding was provided by World Rugby, the Rugby Football Union and Premiership Rugby.

**Data sharing:** Anonymised data available upon reasonable request.

**Ethical approval:** This project was approved by the University's Research Ethics Committee, University of Ulster (#REC-21-0061) and University of Otago Human Ethics Committee (#H21-056). The study was performed in accordance with the standards of ethics outlined in the Declaration of Helsinki.

**Consent to participate:** All participants provided written consent.

**Patient and public involvement:** Patients and/or the public were not involved in the design, conduct, reporting, or dissemination plans of this research.

**Acknowledgements:** The authors would like to thank all staff and players at the participating clubs for their time and involvement in this study. The authors would also like to thank StatsPerform for providing the authors access to their platform. The Rugby Players Association were supportive, endorsed and helped promote the study.

**Abstract**

**Objectives**: To identify a head acceleration event (HAE) severity measure associated with HIA1 removals in elite-level rugby union.

**Methods**: HAEs were recorded from 215 men and 325 women with 30 and 28 HIA1 removals from men and women, respectively. Logistical regression were calculated to identify if peak power, maximum principal strain (MPS) and/or Head Acceleration Response Metric (HARM) were associated with HIA1 events compared to non-cases. Optimal threshold values were determined using the Youden Index. Area under the curve (AUC) were compared using a paired-sample approach. Significant differences were set at p<0.05.

**Results**: All three severity measures were associated with HIA1 removals in both the men's and women's game. Power performed greatest for HIA1 removals in both the men's and women's games, based on overall AUC, sensitivity, and specificity values. HARM and MPS were found to perform lower than PLA in the women's game based on AUC comparisons (p=0.006 and 0.001, respectively), with MPS performing lower than PAA (p=0.001).

**Conclusion**: The findings progress our understanding of HAE measures associated with HIA1 removals. Peak power, a measure based on fundamental mechanics and commonly used in sports performance, may be a suitable HAE severity measure.

**What is already known on this topic**

- In most sports, current suspected concussion detection methods rely on visual identification
- Peak head kinematic values are often used as a proxy for Head Acceleration Event (HAE) severity, though this has led to inconsistencies in the literature.

**What this study adds**

- Peak power may be a suitable HAE severity measure in sport.
- Peak power had the greatest association with Head Injury Assessment (HIA1) removals in men's and women's professional rugby union when compared to other severity measures.

**How this study might affect research, practice or policy**

- Peak power has the potential to be utilised as a severity measure for HAE mitigation strategies and suspected concussion detection tools in sport.
- Peak power may be easier to adopt as a severity measure by players, coaches and other stakeholders owing to its common use in sports performance.

1. Introduction

Identifying suspected concussions on the field remains challenging in sport.(1) In most sports, current detection methods primarily rely on visual identification and video review by sideline medical practitioners, who look for signs such as cognitive and balance abnormalities.(2) If no observable signs of concussion are present, detection depends on player-reported symptoms. In elite rugby union, suspected concussions lead to immediate removal from play for either permanent exclusion or a temporary 12-minute assessment as part of the Head Injury Assessment 1 (HIA1) protocol.(3) The HIA process continues with two post-match evaluations within 2 hours (HIA2) and 36–48 hours (HIA3) using the SCAT6 protocol.(3) Studies indicate that approximately 20% of concussions in elite men's rugby union are not identified on-field, despite video evidence showing signs of concussion at the time.(3)

Head Acceleration Events (HAEs) occur in sport through direct or indirect head loading with more severe events associated with concussion risk.(4) However, it is still unclear what linear and/or rotational head kinematic measures constitute a more severe HAE with peak kinematic values (e.g., Peak Linear Acceleration (PLA), Peak Angular Acceleration (PAA) and Peak Change in Angular Velocity (dPAV)) often used as a proxy.(1)

Instrumented mouthguards (iMGs) have proven effective for measuring head kinematics and are superior to other wearable head sensors (e.g., skin patches) due to a more rigid coupling to the skull.(5) World Rugby has introduced iMGs at the elite level to aid current HIA detection procedures, particularly where players may lack visible signs.(6) PLA and PAA thresholds (75g and 4.5krad/s$^2$ for men and 65g and 4.5krad/s$^2$ for women) are utilised, though these are based on HAE match incidence rather than a direct link to suspected/confirmed concussions.(6) Field-based iMG studies in sport have historically been male-focused and lack suspected/confirmed concussion cases.(1) A recent study found that PLA and dPAV were associated with male HIAs but that PAA was associated with female HIAs.(6) The inconsistency in peak head kinematic measures associated with men's and women's HIA events undermines their potential as an HAE severity measure. The omission of a clear iMG-based severity measure for HAE can lead to ineffectiveness in practice and confusion amongst practitioners/stakeholders, ultimately acting as a barrier to iMG adoption in sport.(7) The aim of this study was to identify an HAE severity measure associated with HIA1 removals in elite-level rugby union.

2. Methods

*2.1.    Study Design*

Data was collected from previously published studies from elite-level Premiership (men), Premier 15s (women) and Farah Palmer Cup (women) competitions utilising the Prevent Biometrics iMG system.(8-10) The iMGs incorporate an accelerometer and gyroscope sampling at 3200Hz with measurement ranges of ±200g and ±35rad/s, respectively. An embedded infrared proximity sensor assesses the iMG's coupling to the upper dentition during HAEs. Previous studies have validated the Prevent Biometrics iMG in both field and laboratory environments.(11-14) The concordance correlation coefficient for peak linear acceleration (PLA) and peak angular acceleration (PAA) measurements ranged between 0.97-

0.98 and 0.91-0.98, respectively, when compared to reference head form measurements.(12,13)

An HAE was identified when linear acceleration at the mouthguard exceeded 8g on a single accelerometer axis.(15) HAE kinematics were recorded 10ms pre-trigger and 40ms post-trigger. For reporting, kinematic signals were transformed to the head's centre of gravity (CG) following SAE J211 standards.(16) A recording threshold of 400rad/s² and 5g at the head CG were set and exhibited a Positive Predictive Value (PPV) of 0.99 (95% CI 0.97–1.00) for identifying contact-related HAEs.(8) For each HAE utilised in the current study, three severity measures were calculated:

### 2.1.1. Head Acceleration Response Metric (HARM)

The Head Acceleration Response Metric (HARM) is currently used as a severity measure to assess American Football helmet performance for the National Football League (NFL).[Bailey] In brief, HARM is a combination of the rotational-based 'Diffuse Axonal Multi-Axis General Evaluation' (DAMAGE) and linear-based 'Head Injury Criterion' (HIC) metrics, see Equation 1.(17,18) The combination of a linear and rotational metric was shown to better distinguish between concussion and non-injurious events in the development of HARM.

$$HARM = C_1 HIC + C_2 DAMAGE \quad [1]$$

where $C_1$ = 0.0148 and $C_2$ = 15.6 are constants determined from fits to head kinematics measured in test dummy reconstructions.

### 2.1.2. Maximum Principal Strain (MPS)

Finite element (FE) brain models are computational tools that examine the mechanical response of the brain at a tissue level to head loading.(19) Previous finite element brain model studies have shown that maximum principal strain (MPS) is the key mechanical metric that predicts concussion and traumatic brain injury.(20-22) An instantaneous brain strain model was utilised to calculate the 95th percentile MPS in the current study.(23)

### 2.1.3. Power

It has been postulated that injury is dependent on the rate at which energy is transferred to the body.(24,25) Accordingly, HAE severity may relate to the maximum value associated with the rate of change of kinetic energy that the head undergoes during a HAE (i.e., peak power), see Equation 2.

$$Peak\ Power = \left[ I_{xx}\ \alpha_x \int \alpha_x\ \partial t + I_{yy}\ \alpha_y \int \alpha_y\ \partial t + I_{zz}\ \alpha_z \int \alpha_z\ \partial t + m a_x \int a_x\ \partial t + m a_y \int a_y\ \partial t + m a_z \int a_z\ \partial t \right]_{max} \quad [2]$$

Where $I_{xx}$, $I_{yy}$, $I_{zz}$ are the componential moments of inertia of the head (kg.m$^2$), $m$ is the head mass (kg), $\partial t$ is the infinitesimal change in time (s), $\alpha_x$, $\alpha_y$, $\alpha_z$ are the componential angular accelerations of the head (rad/s$^2$) and $a_x$, $a_y$, $a_z$ are the componential linear accelerations of the head (m/s$^2$). All head components are in the SAE J211 coordinate system. Since power must be calculated relative to the head reference frame, at time equal zero the velocity associated with power must also equal zero.(24,25) Peak power can be considered synonymous with the measure Head Impact Power.(25) For this study, head mass was approximated based on average male and female cadaveric data (4.1 kg and 3.2 kg, respectively)(26) and moments of inertia based on Equations 3-5.(26) MATLAB code for the calculation of peak power utilised in this study is openly available on GitHub.(27)

$$I_{xx}(kg.cm^2) = 74.8m - 125.5 \qquad [3]$$

$$I_{yy}(kg.cm^2) = 71.4m - 90.2 \qquad [4]$$

$$I_{zz}(kg.cm^2) = 45.6m - 26.5 \qquad [5]$$

### 2.2.     iMG and HIA Event Identification

Removals from play for HIA1 assessments were obtained from the World Rugby SCRM database.(6) The SCRM App securely records all clinical assessments and HIA protocol data globally, incorporating in-built validation checks to enhance data accuracy. An independent researcher undertakes weekly quality control to ensure data accuracy for research purposes.

To identify the HAE event inciting an HIA1 removal, match footage and event data were sourced from StatsPerform (Chicago, Illinois, USA). The match data included details on player contact events (e.g., tackles, carries, rucks) and removal timings. For players removed for HIA1 assessments, the time of removal was used to synchronise iMG HAE timestamps with the contact events.(6) The contact events preceding the player's removal were reviewed to identify the HAE responsible for the HIA1, similar to Allan et al.(6) If the HAE was not clearly identifiable from the video footage, the HIA1 case was excluded from the analysis, and potential HAEs leading to the player's removal were removed.(6) Over the included competitions, match HAEs were recorded from 215 individual men and 325 individual women. A total of 30 and 28 HIA1 removals from 27 and 27 individual players wearing an iMG were identified in the men's and women's cohorts, respectively.

*2.3.    Statistical Analysis*

All statistical analyses were conducted using commercially available software (IBM® SPSS®v.29). Ten random non-case impacts (i.e. HAEs that did not lead to an HIA1 removal) were taken per unique player with ten or more impacts (2150 for men and 3250 for women) to limit oversampling of the non-case events in relation to the HIA1 events.(6) No non-case event was included more than once across the ten random impacts. Simple binary logistical regression and odd ratios (OR) with 95% confidence intervals (CI) were calculated to identify if peak power, MPS and/or HARM were associated with HIA1 events compared to non-cases.(6) Receiver Operator Characteristic curves (ROC) were calculated for the independent variables (Power, MPS, HARM, PLA, PAA, and dPAV) for men and women separately, and optimal thresholds for HIA1 player removal were calculated.(6) Optimal threshold values were determined using the Youden Index, which maximises the independent variables' sensitivity and specificity.(6) Area under the curve (AUC) were compared using the paired-sample approach built into the statistical software. Significant differences were set at $p<0.05$.

3. Results

All three severity measures were associated with HIA1 removals in both the men's and women's game (Table 1). Figure 1 shows the breakdown of the kinematic variables for the HIA 1 and non-cases for both men and women. Power performed greatest from the three severity measures for HIA1 removals in both the men's and women's games, based on overall AUC, sensitivity, and specificity values (Table 2 & 3; Figure 2). Power and HARM performed greater than dPAV in the men's and women's games based on AUC comparisons (Table 4). HARM and MPS were found to perform lower than PLA in the women's game, based on AUC comparisons, with MPS also performing lower than PAA (Table 4).

Table 1. Logistic regression coefficients, OR and p-values for the three severity measures in the men's and women's game.

|  | Coefficients | AUC | Sensitivity | Specificity |
|---|---|---|---|---|
| | | **Men** | | |
| Power | 1.001 (1.001-1.001) | 0.961 (0.924-0.998) | 90.00% | 91.30% |
| MPS | 3.03e16 (1.61e13 - 5.72e19) | 0.948 (0.906-0.990) | 86.70% | 94.50% |
| HARM | 4.206 (3.191 - 5.543) | 0.954 (0.914-0.994) | 86.70% | 95.00% |
| | | **Women** | | |
| Power | 1.001 (1.001-1.001) | 0.923 (0.862-0.983) | 82.10% | 93.70% |
| MPS | 1.47e10 (5.24e7 - 4.11e12) | 0.849 (0.774-0.924) | 82.10% | 76.20% |
| HARM | 3.138 (2.488-3.959) | 0.883 (0.808-0.958) | 71.40% | 94.30% |

Table 2. Median and quartile values for the three severity and kinematic measures with AUC and cut-off value for sensitivity and specificity in the men's game.

|  |  | Median | Q1-Q3 | AUC | Cut-off | Sensitivity | Specificity |
|---|---|---|---|---|---|---|---|
| Power (W) | Non-Case | 427.43 | (230.92-769.67) | 0.961 (0.923-0.998) | 1508.25 | 90.00% | 91.30% |
|  | HIA1 | 6002.07 | (3709.22-8478.88) |  |  |  |  |
| MPS | Non-Case | 0.09 | (0.08-0.12) | 0.948 (0.906-0.991) | 0.17 | 86.70% | 94.50% |
|  | HIA1 | 0.23 | (0.20-0.27) |  |  |  |  |
| HARM | Non-Case | 1.23 | (0.88-1.68) | 0.954 (0.914-0.995) | 2.87 | 86.70% | 95.00% |
|  | HIA1 | 5.41 | (3.85-6.33) |  |  |  |  |
| PAA (krad/s$^2$) | Non-Case | 0.91 | (0.66-1.35) | 0.937 (0.886-0.987) | 1.96 | 86.70% | 89.20% |
|  | HIA1 | 4.07 | (2.60-6.22) |  |  |  |  |
| PLA (g) | Non-Case | 11.42 | (8.37-17.13) | 0.947 (0.906-0.989) | 30.64 | 86.70% | 93.90% |
|  | HIA1 | 56.47 | (34.48-70.59) |  |  |  |  |
| dPAV (rad/s) | Non-Case | 7.99 | (5.53-11.37) | 0.927 (0.875-0.980) | 14.75 | 86.70% | 88.60% |
|  | HIA1 | 23.09 | (18.43-32.26) |  |  |  |  |

Table 3. Median and quartile values for the three severity and kinematic measures with AUC and cut-off value for sensitivity and specificity in the women's game.

| | | Median | Q1-Q3 | AUC | Cut-off | Sensitivity | Specificity |
|---|---|---|---|---|---|---|---|
| Power (W) | Non-Case | 335.21 | (190.12-583.52) | 0.923 (0.861-0.984) | 1193.78 | 82.10% | 93.70% |
| | HIA1 | 2184.62 | (1397.23-4668.55) | | | | |
| MPS | Non-Case | 0.09 | (0.08-0.12) | 0.849 (0.773-0.926) | 0.12 | 82.10% | 76.70% |
| | HIA1 | 0.15 | (0.12-0.22) | | | | |
| HARM | Non-Case | 1.22 | (0.88-1.69) | 0.883 (0.807-0.959) | 2.67 | 71.40% | 94.30% |
| | HIA1 | 2.91 | (1.90-4.96) | | | | |
| PAA (krad/s$^2$) | Non-Case | 0.90 | (0.65-1.33) | 0.917 (0.844-0.990) | 1.68 | 92.90% | 86.50% |
| | HIA1 | 3.17 | (2.07-4.95) | | | | |
| PLA (g) | Non-Case | 10.91 | (8.13-15.51) | 0.947 (0.911-0.983) | 25.05 | 85.70% | 92.80% |
| | HIA1 | 43.13 | (27.62-60.49) | | | | |
| dPAV (rad/s) | Non-Case | 8.14 | (5.61-11.68) | 0.821 (0.738-0.903) | 11.16 | 82.10% | 72.50% |
| | HIA1 | 16.92 | (11.42-20.79) | | | | |

Table 4. Paired-sample AUC comparisons based on the ROC analysis in the men's and women's games. Significant differences are represented by an asterisk (*).

| Test Pairs | AUC Difference | 95% CI Lower Bound | 95% CI Upper Bound | p-value |
|---|---|---|---|---|
| **Men** | | | | |
| Power - MPS | 0.012 | -0.007 | 0.032 | 0.214 |
| Power - HARM | 0.006 | -0.017 | 0.030 | 0.594 |
| Power - PAA | 0.024 | -0.011 | 0.059 | 0.179 |
| Power - PLA | 0.013 | 0.000 | 0.027 | 0.052 |
| Power - dPAV | 0.034 | 0.006 | 0.061 | 0.016* |
| MPS - HARM | -0.006 | -0.015 | 0.003 | 0.167 |
| MPS - PAA | 0.012 | -0.017 | 0.040 | 0.418 |
| MPS - PLA | 0.001 | -0.022 | 0.024 | 0.940 |
| MPS - dPAV | 0.021 | 0.007 | 0.035 | 0.003* |
| HARM - PAA | 0.018 | -0.006 | 0.042 | 0.147 |
| HARM - PLA | 0.007 | -0.018 | 0.032 | 0.581 |
| HARM - dPAV | 0.027 | 0.013 | 0.041 | 0.001* |
| **Women** | | | | |
| Power - MPS | 0.074 | -0.007 | 0.154 | 0.074 |
| Power - HARM | 0.040 | -0.023 | 0.102 | 0.215 |
| Power - PAA | 0.006 | -0.078 | 0.089 | 0.892 |
| Power - PLA | -0.024 | -0.064 | 0.016 | 0.241 |
| Power - dPAV | 0.102 | 0.038 | 0.165 | 0.002* |
| MPS - HARM | -0.034 | -0.078 | 0.011 | 0.135 |
| MPS - PAA | -0.068 | -0.108 | -0.028 | 0.001* |
| MPS - PLA | -0.098 | -0.154 | -0.041 | 0.001* |
| MPS - dPAV | 0.028 | -0.027 | 0.084 | 0.316 |
| HARM - PAA | -0.034 | -0.076 | 0.008 | 0.112 |
| HARM - PLA | -0.064 | -0.110 | -0.018 | 0.006* |
| HARM - dPAV | 0.062 | 0.026 | 0.098 | 0.001* |

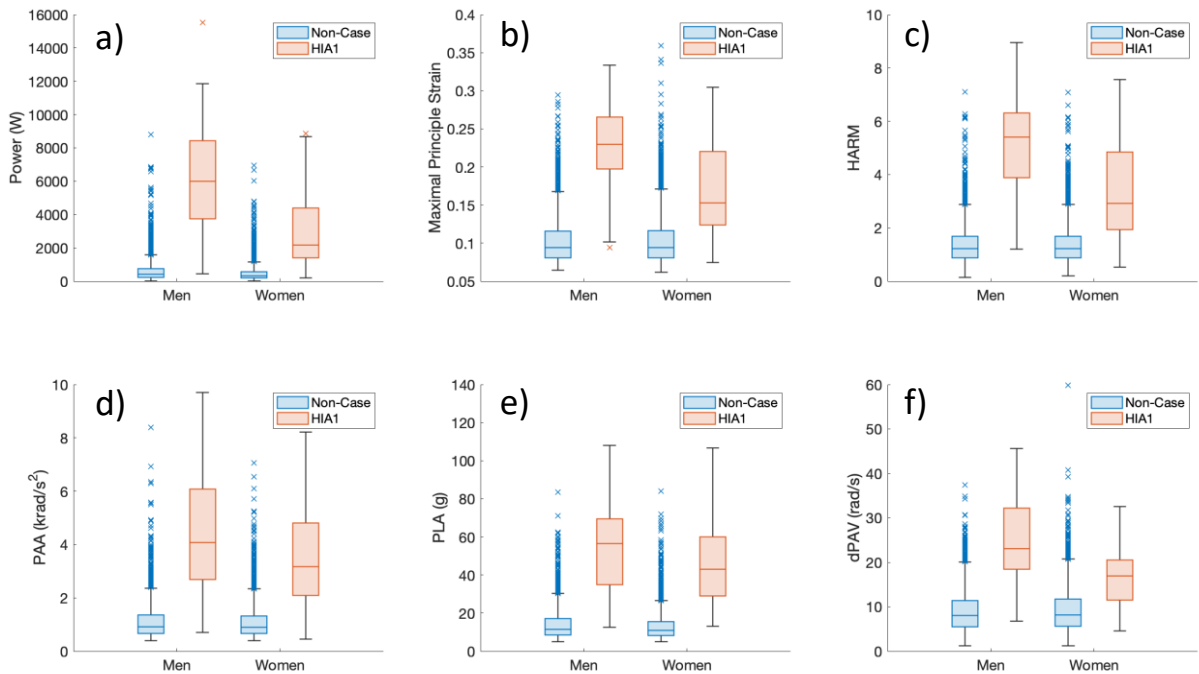

Figure 1. Breakdown of the three severity (a-c) and kinematic measures (d-f) in the men's and women's game illustrating median (box centre line), interquartile range (IQR; box), outliers greater than 1.5 x IQR (crosses) and whiskers (nonoutlier maximum/minimum).

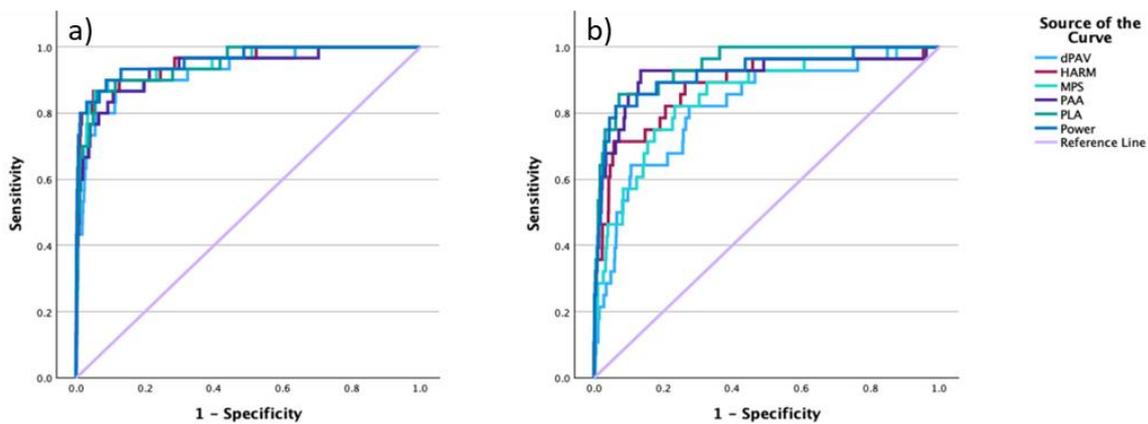

Figure 2. ROC analysis of HIA1 and non-cases for the (a) men's and (b) women's game.

4. Discussion

*4.1.    HAE severity measure*

Peak power appears to be the best performing and most consistent severity measure associated with HIA1 removals during match play in men's and women's professional rugby union. Peak power has the potential to be utilised as a severity measure for research focused on HAE incidence and mechanisms, mitigation strategies and suspected concussion detection tools. The peak power equation is based on fundamental mechanics rather than empirical evidence and includes six degree-of-freedom head acceleration and velocity measures (the latter are represented by the integral terms in Equation 2). Peak power is a common metric already used in sports performance (e.g., strength and conditioning testing)(28) and, therefore, may be easier to adopt as a severity measure by players, coaches and other stakeholders rather than multiple peak kinematic values, which have previously led to confusion.

Peak power, with a six degree-of-freedom head acceleration and velocity measure, performing best in the current study may shed light on conflicting research in the literature that has found different peak kinematic values associated with concussion/suspected concussion. Data from helmet sensor field-based studies have illustrated that rotational acceleration, in particular, is associated with concussion.(29) However, other helmet sensor studies found rotational acceleration to be a significantly worse predictor of concussion than linear acceleration.(29) The purpose of the statistical analysis in the current study is not to derive any form of diagnostic tests, nor to propose HIA1 removal thresholds. Instead, these findings provide a step forward towards understanding HAE severity and what measures may be associated with HIA1 removal. iMGs are not currently a replacement for the HIA process in rugby union but an additional tool to aid clinical decision making for HIA removals. Removals based on peak power threshold values should be assessed formally to ensure high performance in terms of sensitivity, specificity and other accuracy measures. For example, a high rate of false positive cases could overwhelm medical support staff and disrupt matches to an extent that iMG use is rejected by coaches and players.(6) In the women's game, HARM and MPS underperformed relative to certain peak head kinematics, potentially highlighting the need for HAE severity measures to be sex specific/adaptable.(1)

*4.2.    Limitations*

High severity measures were identified in non-clinical cases (Figure 1), although no real-time observations of clinical signs, symptoms, or behavioural changes were made. These signs may have been absent or the player may have continued to play without disclosing or displaying any effects of the HAE.(3) It remains unclear whether these HAEs resulted in the clinical presentation of signs and symptoms post-match. Analysis of these cases should be a focus of future work.

The current study may not comprehensively capture the range of playing styles and conditions across all levels of rugby globally. HAE severity measures could vary in different rugby cohorts, especially in youth, as well as amateur-level games.

Kinematic signal processing was performed using the Prevent Biometrics system, similar to other commercially available iMG systems.(12) The kinematic signal processing used in this study has been included in validation studies for the Prevent Biometrics iMG system,(12) and is currently utilised in professional rugby.(6) However, a standardised and openly available signal processing method for iMG systems, such as the HEADSport filter, may be necessary.(30) A consensus-agreed and consistent signal processing approach is crucial for enabling inter-study comparisons within and between different sports, particularly when multiple iMG systems are utilised.[17]

The MPS measures in the current study were based on a validated instantaneous brain strain estimation model trained on a large number finite element brain model predictions.(23) The rationale for the selection was that an instantaneous brain strain measure would be practically required pitch-side for HIA detection. Finite element and other biomechanical modelling can complement iMG data in uncovering injury mechanisms.(1,31) The head mass, and thus moment of inertia were approximated for the peak power calculation. However, a more subject-specific approach could be beneficial by measuring head circumference (C) and utilising Equation 6.(26)

$$Mass\ (kg) = 0.23C\ (cm) - 9.33 \qquad [6]$$

In future research, the incorporation of clinical outcomes from the entire HIA process will allow for an evaluation of the diagnostic accuracy of iMG in concussion detection. However, the current mandate by World Rugby is to use iMG as part of the criteria for identifying players who require the HIA1 screen, rather than for direct concussion diagnosis. This approach facilitates a larger sample size for evaluation, and in the future, a combined approach could investigate the associations between HAE severity, HIA1 indicators, and concussion outcomes.

5. Conclusion

Peak power, a measure based on fundamental mechanics, may be a suitable HAE severity measure in sport. Peak power was most consistently associated with HIA1 removals during match play in men's and women's professional rugby union. All three severity measures were associated with HIA1 removals in both the men's and women's game. However, peak power performed greatest for HIA1 removals in men's and women's professional rugby union, based on overall AUC, sensitivity, and specificity values. Power and HARM performed greater than dPAV in the men's and women's game based on AUC comparisons. HARM and MPS were found to perform lower than PLA in the women's game, based on AUC comparisons, with MPS also performing lower than PAA. The findings progress our understanding of HAE severity and measures associated with HIA1 removals. Peak power may be easier to adopt as a severity measure by players, coaches and other stakeholders owing to its common use in sports performance.

6. Policy Implications

Peak power has the potential to be utilised as a severity measure for HAE mitigation strategies and suspected concussion detection tools in sport.